\documentclass[12pt]{article}

\usepackage{amsfonts}

\usepackage{amssymb}
\newtheorem{thm}{Theorem}

\newtheorem{cor}[thm]{Corollary}

\newtheorem{lemma}[thm]{Lemma}

\newtheorem{prop}[thm]{Proposition}

\newtheorem{defn}[thm]{Definition}

\def\pf{\medbreak\noindent{\bf Proof:}\enspace}

\def\tr{\hbox{Tr}}

\def\be{\begin{eqnarray}}

\def\ee{\end{eqnarray}}

\def\bee{\begin{eqnarray*}}

\def\eee{\end{eqnarray*}}

\def\ts{\textstyle}

\def\bra{\langle}

\def\ket{\rangle}

\def\kb{ \ket \bra }

\def\rt2{\ts \frac{1}{\sqrt{2}} }

\def\nrm{\vert \vert}

\title{
Distinguishing Bipartitite Orthogonal States using
LOCC: Best and Worst Cases}

\author{Michael
Nathanson
\\
\\ Department of Mathematics
\\ Northeastern University
\\ Boston MA 02115
\\
{\normalsize nathanson.m@neu.edu}
}
\begin{document}

\maketitle

\begin{abstract}
Two types of results are presented for distinguishing pure bipartite quantum states using
Local
Operations and Classical Communications. We examine
sets of states that can be perfectly distinguished, in
particular showing that any three orthogonal maximally
entangled states in $C^3 \otimes C^3$ form such a set.
In cases where orthogonal
states cannot be distinguished, we obtain
upper bounds for the probability of error using LOCC
taken over all sets of $k$ orthogonal states in $C^n
\otimes C^m$.Ê In the process of proving these bounds,
we identify some sets of orthogonal states for which
perfect distinguishability is not possible.

\end{abstract}

%\tableofcontents

%\bigskip

\section{Introduction}
There is much interest in understanding what can and
cannot be achieved using Local Operations and
Classical
Communications 
(LOCC) on a composite quantum system, pursued with an
eye towards applications in communication and
cryptography. One of the first and most
basic problems in LOCC is that of distinguishing
orthogonal quantum states. While some direct
applications of this problem do exist (for instance,
data hiding \cite{TDL} and corrected channels \cite{GW,HK}), these are limited by the
usual assumption that no
additional entanglement exists between the two
parties. However, the problem of LOCC discrimination has proved a fertile area for attempts to better
understand the
relationship between entanglement and locality, the
mysterious interplay that underlies virtually all
quantum communication and cryptography protocols. It
is in this spirit that the current work is undertaken.

The set-up for bipartite LOCC is quite simple: Two
parties (by convention Alice and Bob) are physically
separate but share a quantum state. Each may perform
local quantum operations on his/her piece of the
system, but the two may only communicate through a
classical channel. In this paper, we will suppose that
Alice and Bob share one of a known set of orthogonal
states; their task is to determine the identity of
this state (even if it is destroyed in the process).
Since the possible states are orthogonal, they clearly could
be distinguished and preserved were global
operations permitted. 

The most
fundamental and surprising results in this area are
those of Walgate, et al.,\cite{WSHV} that
\textit{any} two orthogonal states can always be locally
distinguished; and of Bennett, et al.,\cite{QNWE}
that there exists a basis of product states that
cannot be distinguished with LOCC. These two facts
demonstrate that there is no simple relationship
between entanglement and locality, which has led to
further
exploration, e.g. \cite{HSSH,WH}.Ê

Following the definitive result for two states
\cite{WSHV}, work has been done to identify larger
sets of orthogonal states that can and cannot be
perfectly distinguished with LOCC. Both \cite{HF} and
\cite{GKRS} looked at generalized Bell bases in $C^n
\otimes C^n$. Fan \cite{HF} showed that any $k$ such
states can be perfectly distinguished if $n$ is prime
and $k(k-1) \le 2n$, in particular in the case
$k=n=3$. 
The question was posed in 
\cite{GKRS}
whether \textit{any} 3 maximally entangled states
could be
distinguished; we answer this
question in the affirmative. We also
also give a sufficient condition for perfect
distinguishability among maximally entangled states in
$C^n \otimes C^n$ using unbiased bases, thus providing an
alternative proof of the result in \cite{HF}.

It not always possible to
perfectly distinguish $k$ orthogonal vectors when $k >
2$. For instance, Ghosh, et al., showed thatÊ $k$
generalized Bell states in $C^n \otimes C^n$ 
cannot be distinguished with LOCC if $k >n$. \cite{GKRS, GKRSS}
The second part of this paper
establishes lower bounds on the effectiveness
of probabilistic LOCC discrimination of orthogonal vectors. If Alice
and Bob share one of $k$
arbitrary
orthogonal vectors in $C^n \otimes C^n$, 
what is their guaranteed minimal probability of
correctly identifying it? And which sets of states
achieve this minimum? These questions have an
immediate application to a data hiding set-up as described
in \cite{TDL}, in which a `Boss' can clear prior
entanglement between Alice and Bob before giving them
pieces of a secret quantum state to work on. 

It is
shown that for $2 \le k \le 4$, 
$k$ arbitrary orthogonal vectors in $C^m \otimes C^n$
can be correctly identified with probability at least
$\frac{2}{k}$, and this bound is tight. An interesting
fact is that this
does not
depend on the dimension of the overall space--the
worst case occurs when the states are embedded in a
$C^2 \otimes C^2$ subspace. Our final result
translates these ideas into the more familiar language
of mutual information and recovers a bound implied by
\cite{BHSS}.

The bounds from these propositions identify sets
of vectors for which perfect distinguishability is
impossible. In particular, we generalize
\cite{GKRS} to show that no $k$ maximally entangled
states can be perfectly distinguished if $k > n$. The
bounds also lead to the well-known result of
Horodecki, et
al., \cite{HSSH} that a 
complete basis of perfectly distinguishable vectors
must be a product basis. 

As a final comment, we note the distinction made in
\cite{SDeR} between LOCC protocols that have so-called
infinite resources and those that use a finite number
of rounds of communication and remain in
finite-dimensional ancillary spaces. The results in
this paper are established under the assumption that
all protocols terminate with
probability one and that each ancillary system is
finite dimensional.

The paper is organized as follows: Section
\ref{statement} states the results and gives
necessary background, and Sections \ref{perfect}
and \ref{imperfect} provide the proofs.

\section{Statement of Results}\label{statement}

Following the result \cite{WSHV},Ê we
would like to identify sets of $k$ orthogonal vectors
that can
be perfectly distinguished with LOCC for $k > 2$.Ê For
instance, it is immediate that any three
orthogonal states can be perfectly
distinguished if two of them are product states. Also,
Êfrom $\cite{HF}$, any $3$ states
of a generalized Bell basis of $C^n \otimes C^n$ can
be distinguished if $n \ge 3$;Ê the question for
general maximally entangled vectors in $C^3 \otimes
C^3$ is noted but not
answered in \cite{GKRS}.

\begin{prop}\label{3C3} Any three orthogonal maximally
entangled
states in $C^3 \otimes C^3$ can be perfectly
distinguished using LOCC. 

\end{prop}

It is not clear whether any 3 orthogonal maximally
entangled states are distinguishable in $C^n \otimes
C^n$. However, the following proposition gives a
sufficient condition
for distinguishing maximally entangled states using
the idea of mutually unbiased bases, which arise in
several area of quantum information (see, for instance
\cite{PR,W}). The more general notion of a
common unbiased basis is not well-studied but is 
defined here for convenience:
\begin{defn}
Let ${\cal{A}} = \lbrace {\cal{A}}_i : i \in {\cal{I}}
\rbrace$ be a family of
orthonormal bases of $C^n$, withÊ $ {\cal{A}}_i =
\lbrace \vert a_{i1} \ket, \vert a_{i2} \ket, \ldots,
\vert a_{in} \ket \rbrace$ and ${\cal{I}}$ some
indexing set. 

A basis ${\cal{B}}$ of $C^n$ is a \textit{common
unbiased
basis} for ${\cal{A}}$ if, for all $\vert b \ket \in
{\cal{B}}$ and for all $i \in {\cal{I}},1 \le j \le
n$:
\be
\vert \bra b \vert a_{ij} \ket \vert^2 = \frac{1}{n}
\ee
\end{defn}

So, a set of bases ${\cal{A}}$ is mutually unbiased if
and only if for all $i \inÊ {\cal{I}}$, ${\cal{A}}_i$ is aÊ common unbiased basis
forÊ ${\cal{A}} - \lbraceÊ {\cal{A}}_i\rbrace$. 

In the following proposition, we write our states in
terms of a (non-canonical) standard maximally
entangled state of $C^n \otimes C^n$:
\be
\vert ME_n \ket & := & \frac{1}{\sqrt{n}}\label{MEn}
\sum_{j=0}^{n-1} \vert j
\ket \vert j \ket 
\ee

\begin{prop}\label{CUB}
Let $\vert \Psi_1 \ket, \vert \Psi_2 \ket, \ldots
\vert
\Psi_k \ket$ be orthogonal, maximally entangled
vectors in $C^n\otimes C^n$, with $\vert \Psi_i \ket =
(I \otimes B_i)\vert ME_n
\ket$.

For each pair $(i,j)$, let ${\cal{A}} _{ij}$ be a
basis of eigenvectors of $B_i^\dagger B_j$, and letÊ 
\bee
{\cal{A}}Ê = \lbrace {\cal{A}} _{ij}: 1 \le i <Ê j \le
k \rbrace
\eee

If the family ${\cal{A}}$ has a common unbiased basis,
then the $k$ states
can be perfectly distinguished by LOCC.
\end{prop}

The result is actually more general--we
do not require that the states be maximally
entangled, only that the matrices $B_i^\dagger B_j$ 
be
diagonalizable. For instance, we could use the same proof to
show that any simultaneously diagonalizable orthogonal
states can be locally distinguished. These are sets of the form
\be
\lbrace \vert \varphi_i \ket = \sum_{j=0}^{n-1} u_{ij} \vert jj\ket,Ê 1 \le i \le n \rbrace
\ee
where $u$ is an $n \times n$ unitary matrix. 

The main result of \cite{HF}Ê follows from
Proposition \ref{CUB} . It involves the generalized
Pauli
matrices $Z = \sum_jÊ e^{2\pi i j/ n} \vert j \kb j
\vert$ and
$X = \sum_j \vert j \kb j+1 \vert$ and the generalized
Bell basis 
\be \label{bellbasis}
BB_n := \lbrace (I \otimes X^mZ^l)\vert ME_n \ket : 0
\le m,l \le n-1 \rbrace \subset C^n \otimes C^n
\ee

\begin{cor} (H. Fan)
Let $\vert \Psi_1 \ket, \vert \Psi_2 \ket, \ldots
\vert
\Psi_k \ket$ be orthogonal, maximally entangled
vectors in $C^n\otimes C^n$, with $n$ prime and $\vert
\Psi_i \ketÊ \in BB_n$. 

Then if $k(k-1)/2Ê \le n$, the $k$ vectors
can be perfectly distinguished by LOCC. 
\end{cor}

\pf
This follows from the fact that for $n$
prime, the eigenbases of $\lbrace X^mZ^l : 0 \le l,m <
n \rbrace$ form a maximum set
of $(n+1)$ mutually unbiased bases in $C^n$.\cite{PR}
Up to a
global
phase,
\be
(X^{m_i}Z^{l_i})^\dagger (X^{m_j}Z^{l_j}) \equiv 
X^{m_j -
m_i}Z^{l_j - l_i}
\ee
so the eigenbases of the pairwise products also belong
to the set of
mutually unbiased bases. As long as the number of
pairs $(i,j)$ is less than
the number of mutually unbiased bases, then there
exists a common unbiased basis and the proposition can
be applied. But
this is the condition that $k(k-1)/2 < n +1 $.

\bigskip

It is not always possible to distinguish maximally
entangled states (\cite{GKRSS}), which raises the
question of how bad it can be (or conversely, what
minimal level
of success is guaranteed). When perfect discrimination is not possible, one possible strategy is unambiguous discrimination, in which either the correct identity of the state is discovered or else a generic error message is returned. Another strategy is minimum error discrimination, in which the protocol always produces one of the possible states but this identification might be incorrect. The challenge in this case is to find a protocol that minimizes the probability of error. It is this problem of minimum error discrimination that we will consider throughout the rest of the paper. 

Suppose Alice and Bob share one of the orthogonal vectors
$\lbrace \vert \Psi_i \ket \rbrace$ with a priori
probabilities $ \lbrace
p_i \rbrace$. They apply an LOCC protocol, which
produces a best guess as to the identity of their
state. Define $P(\lbrace \vert \Psi_i \ket
\rbrace, \lbrace
p_i \rbrace)$ as the probability that Alice and Bob
correctly identify which vector they share,
assuming an optimal strategy is used.  We are
interested in the worst case scenario--what ensembles
of $k$ orthogonal vectors are hardest to distinguish using LOCC?
Initially, we restrict ourselves to maximally
entangled states and define
\be \label{definefme}
f_{me}(k,n) := \min_{\lbrace \vert \Psi_i \ket
\rbrace,
\lbrace p_i \rbrace}Ê P(\lbrace \vert \Psi_i \ket
\rbrace, \lbrace p_i \rbrace)
\ee
where the minimum is taken over probability distributions
$p_i$ and sets of orthogonal maximally entangled
states $\lbrace \vert
\Psi_1
\ket, \ldots, \vert \Psi_k
\ket\rbrace \subset C^n \otimes C^n$. 

We immediately observe that $f_{me}$ is a
nonincreasing
function in both $k$ and $n$; as $k$ and $n$ increase,
the minimum is taken over largerÊ nested
sets. We note that for all $n$, $f_{me}(2,n) =
1$, since
two orthogonal states can always be
distinguished
by
LOCC.\cite{WSHV}Ê Proposition \ref{3C3} is equivalent
to
the fact that $f_{me}(3,3) = 1$.

But there are limitations to what can be done if the
number of vectors is bigger than the dimension:
\begin{prop}\label{fmeprops}
For all $2 \le n \leÊ k \le n^2$,
\be
\frac{2}{k} \le f_{me}(k,n) \le \frac{n}{k}
\ee
In the case $n = 3 \le k \le 9$,
\be
f_{me}(k,3) = \frac{3}{k}
\ee
\end{prop}

We can also define a more general function in which we
remove the assumption that the states are maximally
entangled
\be \label{definef}
f(k,n) := \min_{\lbrace \vert \Psi_i \ket \rbrace,
\lbrace p_i \rbrace}Ê P(\lbrace \vert \Psi_i \ket
\rbrace, \lbrace p_i \rbrace)
\ee
where the minimum is taken over probability distributions
$p_i$ and \textit{all} sets of orthogonal states $\lbrace \vert
\Psi_1
\ket, \ldots, \vert \Psi_k
\ket\rbrace \subset C^n \otimes C^n$. 

Again, $f$ is nonincreasing with respect to $n$
and $k$ and $f(2,n) = 1$. Also, for $k \le m^2 \le
n^2$, $k$ maximally entangled vectors in $C^m \otimes
C^m$ can be embedded in $C^n \otimes C^n$, soÊ $f(k,n)
\le f_{me}(k,m)$.Ê 
The previous results for $f_{me}$ imply bounds on $f$:
\begin{prop}\label{fprops}
For $2 \le n \le k \le n^2$,
\be
\frac{2}{k} \le f(k,n) \le \frac{\lceil \sqrt{k}
\rceil}{k}
\ee
In particular,
\be
f(3,n) = \frac{2}{3} \quad
Ê f(4,n) = \frac{1}{2}
Ê \ee
Ê \end{prop}

The function $f(k,n)$ is defined only when the two
spaces have the same dimension. We could just as
easily have defined $f(k,m,n)$ for $k$ vectors in $C^m
\otimes C^n$ and applied Lemma \ref{upperboundlemma}
to that. However, we have discovered no bounds on this
that don't follow from inclusion; that is, for $m \le
n$, the best we can say is:
\be
f(k,n) &\le f(k,m,n) \le &f(k,m)ÊÊ \quad\quad
k \le m^2 \\
f(k,n) &\le f(k,m,n) \le& \frac{n}{k} \quad \quad m^2
< k
\le mn
\ee

We note that for $k \le 4$, $f(k,n)$
is independent of $n$; the $k$ vectors
are most difficult to distinguish when they are
squeezed into the smallest possible space. It seems
entirely possible that $f(k,n)$ will remain
independent of $n$ even for higher values of $k$. 

Propositions \ref{fmeprops} and \ref{fprops} are
proved using the following lemmas. In fact, most of
the work goes into the proof of Lemma
\ref{upperboundlemma}, as it requires us to analyze
Alice and Bob's protocol in detail. 
\begin{lemma}\label{propertieslemma}
For all $2 \le j \le k \le n^2$,
\be
\frac{j}{k} f_{me}(j,n) &\le& f_{me}(k,n) \\
\frac{j}{k} f(j,n) &\le& f(k,n)
\ee
\end{lemma}
\begin{lemma}\label{upperboundlemma}
Given $k$ equally probable vectors $\lbrace \vert
\Psi_1
\ket, \ldots, \vert \Psi_k \ket \rbrace \subset C^m
\otimes C^n$, $n \le k \le mn$, with the property that
for each $i$,
$\vert \Psi_i \ket = (I \otimes U_i) \vert \Psi_1
\ket$ for $U_i$ unitary. Then the $k$ vectors can be
distinguished using LOCC with probability at most
$\frac{n}{k}$.
\end{lemma}
The assumption in Lemma \ref{upperboundlemma} is
equivalent to the fact that the $C^n$ party can unilaterally
transform $\vert \Psi_i \ket$ intoÊ $\vert \Psi_j
\ket$ for any $i,j$. The lemma includes the special
case in which all the states are maximally entangled.
Also, note that there is no assumption here that the
states are orthogonal, though this is clearly the most
interesting case. 

\textbf{Examples:} Given a basis of 4 orthogonal
maximally entangled
states in $C^2 \otimes C^2$. One naive notion is
ignore two of the possible states and perfectly
distinguish the remaining two, thus achieving the
lower bound in Lemma \ref{propertieslemma}. Lemma
\ref{upperboundlemma}
states that this, in fact, is an optimal strategy for
identifying the given state. Proposition \ref{fprops}
combines the lemmas to say that
this is the worst case for trying to distinguish 4
orthogonal states.

Likewise, given $k>3$ orthogonal maximally entangled
states in $C^3 \otimes C^3$, one can discard
all but three of them and then perfectly distinguish
those that remainÊ using Proposition \ref{3C3}.
Again, the lemma states that
this
is optimal. However, forÊ $k = 4$ or $k=5$, this
succeeds with
probability greater than $\frac{1}{2}$ and so is no
longer the worst case in $C^3 \otimes C^3$. A worse
case would be 4 equally probable maximally entangled
states in a $C^2 \otimes C^2$ subspace. 
Ê 

Finally, we look at an exampleÊ 
using the generalized Bell basis $BB_n$ defined in
(\ref{bellbasis}). Suppose we wish to distinguish the
states in a set $T
\subset BB_n$ withÊ $\vert T \vert = k$. If $n$ is
prime,
then the argument in \cite{HF} implies that Alice and
Bob can correctly
identify
their vectors with probability $\frac{n}{k}$; Lemma
\ref{upperboundlemma} shows that this is in fact
optimal.

The following modification of Lemma
\ref{upperboundlemma} establishes a necessary
condition toÊ distinguish a set of
states:\begin{prop}\label{schmidtprop}
Given $k$ equally probable vectors $\lbrace \vert
\Psi_1
\ket, \ldots, \vert \Psi_k \ket \rbrace \subset C^m
\otimes C^n$ and let $\lambda_M$ be the largest
Schmidt coefficient in any of the $\vert \Psi_i \ket$.
Then the $k$ vectors can be
distinguished using LOCC with probability at most
$\frac{\lambda_M mn}{k}$. 

In particular, if $k$ vectors can be perfectly
distinguished with LOCC, then $\lambda_M \ge
\frac{k}{mn}$.

\end{prop}

It is interesting to note that in
the case of perfect distinguishability, this
proposition gives
a lower bound on the maximal Schmidt coefficient,
while the
result of Chen and Li \cite{CL} gives an upper bound
on the number of nonzero Schmidt coefficients.Ê 

The following generalizes the work of 
\cite{GKRS} by setting $\lambda_M = \frac{1}{n}$
above. 

\begin{cor}
No $k$ maximally entangled
states in $C^n \otimes C^n$ can be perfectly
distinguished with LOCC if $k > n$.
\end{cor}

Both
Proposition \ref{schmidtprop} and the result \cite{CL}
imply the
fundamental result of Horodecki, et al., that a
distinguishable basis must be a product basis
\cite{HSSH}:

\begin{cor} (Horodecki, et al.)
Let $\lbrace \vert \Psi_1 \ket, \ldots,\vert 
\Psi_{mn} \ket \rbrace$ be an orthonormal basis of
$C^m \otimes C^n$, and suppose these vectors can be
perfectly distinguished using LOCC. Then each of the
vectors is a product vector. 
\end{cor}
To see this as a consequence of Proposition
\ref{schmidtprop}, suppose we have have one of the
$\vert
\Psi_i \ket$ with equal probability. Then clearly
$\lambda_M =k / mn= 1$.Ê Examining the proof of
Proposition \ref{schmidtprop} reveals that if $\vert
\Psi_i \ket$ has maximal Schmidt coefficient
$\lambda_i < \lambda_M$, then either $P(Z = i) = 0$ or
else the inequality on $P(Z=V)$ is strict. Neither of
these is
possible with perfect distinguishability, which means
$\lambda_i = \lambda_M =1$ and $\vert \Psi_i \ket$ is
a product state for all $i$.

ÊThese types of results are useful in that they allow
us to identify classes of sets of $k$
vectors in $C^m \otimes C^n$ that cannot be perfectly
distinguished. Also, they provide an upper
bound on the probabilities and allows us to deduce
optimal strategies for correct identification. 

Ê

The function $f(k,n)$ is one way of assessing how much
information Alice and Bob can gain from LOCC
measurements on their vectors. Another
approach would be to use the classical mutual
information
between the identity $V$ of the vector sent and the
outcomes of Alice
and Bob's measurements. (This idea was explored, for
instance, with reference to the
specific 9-state ensemble in \cite{QNWE}.) Let $Y$
represent the
outcomes of the first $r-1$ measurements and $Z$
indicate the final measurement, i.e. the conclusion
as to the value of $V$, and write
\be
I(V; YZ) = H(V) - H(V \vert YZ)
\ee
where $H$ is the Shannon entropy.

As we defined $f(k,n)$, we define a function
$g(k,n)$ based on mutual information. Assuming that
Alice and Bob use optimal measurements, we can
consider $I(V; YZ)$ to be the optimal mutual
information between the input vector $V$ and the
measurement results.
Ê\be
g(k,n) := \min_{\lbrace \vert \Psi_i \ket \rbrace
} 
I(V ; YZ)
\ee
Note that we now assume that all the $k$ vectors are
equally likely; there is no sensible lower bound if
the entropy of the a priori probability distribution
is allowed to approach
zero. 
\begin{prop}\label{infoprop}
The function $g(k,n)$ defined above for $1 < k \le
n^2$ satisfies the following bounds:
\be
\frac{2}{k}\log{2} \le g(k,n) \le \log{\lceil \sqrt{k}
\rceil} 
\ee

\end{prop}
This proposition is proved as a consequence of Lemma
\ref{upperboundlemma}. The same upper bound can be seen as a
consequence of the following inequality given in
\cite{BHSS}:
\be
I_{acc} ^{LOCC}\le S(\rho_A) + S(\rho_B) - \sum_i p_i
S(\rho_A^i)
\ee
where $I_{acc}^{LOCC}$ is the classical mutual accessible
information using LOCC, $S$ is von Neumann entropy,
$\rho = \sum p_i \vert \Psi_i
\kb \Psi_i \vert$, and $\rho_A$ and $\rho_B$ are the
partial traces. 

Let the $\vert
\Psi_i \ket$ be maximally entangled states in $C^n
\otimes C^n$. Then \be
  \rho_A^i &=& \rho_A = \rho_B = \frac{1}{n} I_n ~~ \forall ~ i \\
I_{acc} ^{LOCC}& \le &Ê S(\rho_A) +
S(\rho_B) - \sum_i p_i S(\tr_A( \vert \Psi_i \kb
\Psi_i \vert)) \\
& \le &\log n + \log nÊ - \sum_i p_i S(\tr_A( \vert
\Psi_i \kb \Psi_i \vert)) = \log nÊ \ee
Ê
This gives another way to see that $k$ maximally
entangled states in $C^n \otimes C^n$ cannot be
distinguished if $k > n$.

\textbf{Example:} Recall the set $BB_n$ defined in
(\ref{bellbasis}); it is a generalized Bell basis for
$C^n \otimes C^n$. Suppose Alice and Bob share a
state $\vert \Psi \ket = (I \otimes X^m Z^l )\vert
ME_n
\ket$, uniformly chosen from $BB_n$. Each
measures in the standard basis, allowing them to
perfectly determine the value of $m$ but giving no
information about $l$.

If at this point, they make a guess as to the value of
$l$, they will be correct with probabilty
$\frac{1}{n}$, which saturates the inequality in Lemma
\ref{upperboundlemma}, and hence is optimal for $P(Z =
V)$.

Perhaps more surprising, this protocol is also optimal
with respect to classical mutual information, as 
Ê$I(V;
YZ) = \log n$ and the proof of the upper bound in
Proposition
\ref{infoprop} shows that this is maximal.Ê 

\section{Proofs of Propositions for Distinguishing
Maximally Entangled
States}\label{perfect}
Ê\subsection{Preliminaries}\label{prelims}
As has been previously noted (for instance in
\cite{R}), there is one-to-one correspondance between
statesÊ $\vert \Psi \ket
\in C^n \otimes C^m$ and $m \times n$ complex matrices
$B$ given by $\vert \Psi \ket
Ê= (I \otimes B) \vert ME_n \ket$, where $\vert ME_n
\ket$
isÊ the standard maximally entangled $C^n \otimes C^n$ state defined in
(\ref{MEn}).
Throughout the paper, we
will use the following property,
which was noted in \cite{R} and implicitly used in
\cite{WSHV}:
\begin{lemma}\label{transposelemma}
For any
$m \times n$ matrix $A$ written in the standard basis,
\be
\sqrt{n}(I \otimes A) \vert ME_n \ket = \sqrt{m}(A^T
\otimes I) \vert 
ME_m
\ket
\ee
In particular, setting $m = 1$,
\be
\sqrt{n}(I \otimes \bra v \vert ) \vert ME_n \ket =
\vert \overline{v} \ket
\otimes I \ee
\end{lemma}
where $\vert \overline{v} \ket$ denotes the entrywise complex conjugate of $\vert v \ket$ in the standard basis. 

We adopt the convention of associating 
states $\vert \Psi \ket$ with
$\bra \Psi \vert \Psi \ket = 1$ and $m \times n$ matrices $B$ with $\tr B^\dagger B = n$. This
correspondance has the following immediate
properties: 
\begin{enumerate}
\item{} If $\vert \Psi_i \ket = (I \otimes B_i)\vert
ME_n \ket$ for $i = 1,2$, then $\bra \Psi_1 \vert
\Psi_2
\ket = \frac{1}{n} \tr B_1^\dagger B_2$
\item{} $\nrm B^\dagger B \nrm_\infty =
n\lambda_{M}$,
where $\lambda_{M}$ is the largest Schmidt
coefficient of $\vert \Psi \ket$.
\item{} $\vert \Psi \ket = (I \otimes B)\vert ME_n
\ket \in C^n \otimes 
C^n$
is maximally entangled if and only if $B$ is unitary.
\end{enumerate}
We will use this correspondance throughout what
follows. 
Ê
\subsection{Proof of Proposition \ref{3C3}} 
For $i = 1,2,3,$ write $\vert \Psi_i \ket = (I
\otimes B_i)\vert ME_3
\ket$ withÊ $B_i$ unitary and $\tr B_i^\dagger B_j =
3\delta_{ij}$.Ê The matrix $B_2^\dagger B_1$ is a
traceless
$3\times3$
unitary matrix, so its eigenvalues are $\lbrace 1,
\omega, \omega^2 \rbrace$, with $\omega =
e^{i2\pi/3}$.
The same is also true for $B_3^\dagger B_2$. We write
these matrices in terms of their eigenvectors:
\be
B_2^\dagger B_1 = \sum_{i = 0}^2 \omega^i \vert e_i
\kb e_i
\vert \quad B_3^\dagger B_2 = \sum_{i = 0}^2 \omega^i
\vert
f_i \kb f_i
\vert
\ee
Given $\vert \Psi_i \ket$, for $i$ unknown, 
choose a unitary $U$ and
measure the first system in the basis $\lbrace
\overline{U}
\vert j \ket: j = 0,1,2 \rbrace$, where $\overline{U}$
indicates the entrywise complex conjugate of $U$. If
the outcome of the measurement is $x \in \lbrace 0,1,2
\rbrace$, then Lemma \ref{transposelemma} implies the
state now looks like: 
\be
(\overline{U}\vert x \kb x \vert U^T \otimes I)\vert \Psi_i \ket & = & (\overline{U}\vert x \kb x \vert U^T \otimes B_i)\vert ME_n \ket \\ 
& = & (\overline{U}\vert x \ket \otimes B_i)(\bra x
\vert U^T \otimes I) \vert ME_n \ket \\ 
& = &Ê \frac{1}{\sqrt{n}} (\overline{U}\vert x \ket
\otimes B_i)(I \otimes U \vert x \ket )\\ 
& = & \frac{1}{\sqrt{n}}Ê \overline{U}\vert x \ket
\otimes B_iU\vert x \ket
\ee
In particular, after normalization, the second system
is in the state
\be
B_iU\vert x\ket
\ee
We want to show that for appropriate choice of $U$,
the vectors $\lbrace B_1U\vert x \ket,B_2U\vert x
\ket,B_3U\vert x \ket \rbrace$ are orthogonal for all
$x$. The proof is constructive and is
achieved in 3 steps:
\begin{enumerate}
\item{} Observe that the quantity $\vert \bra
e_i
\vert f_j \ket \vert^2$ depends only on $(j-i)$ mod 3.
\item{} Show that we can adjust the phases of the
$\vert e_i \ket$ and $\vert f_j \ket$ so that we may
assume that $\bra
e_i
\vert f_j \ket $ depends only on $(j-i)$ mod 3.
\item{} Let our unitary $U$ beÊ the Fourier matrix in
the basis $\lbrace \vert e_i \ket \rbrace$ and show
that the vectors $\lbrace B_1U\vert x \ket,B_2U\vert x
\ket,B_3U\vert x \ket \rbrace$ are orthogonal for all
$x$. 
\end{enumerate}
Ê
The proof of each step is given below. Note that all operations on indices are assumed to be taken modulo 3. 
\begin{enumerate}
\item{} Since $\tr B_3^\dagger B_1 = 0$:
\be
0 = \tr B_3^\dagger B_2B_2^\dagger B_1 = \sum_{i,j}
\omega^{i+j}
\vert \bra
e_i \vert f_j \ket \vert^2 = \sum_{i,k} \omega^{k}
\vert \bra e_i \vert
f_{k-i} \ket \vert^2
\ee
For any $a_k \ge 0$, $\sum_{k=0}^2 \omega^k
a_k
= 0$ implies that all the $a_k$ are the same.
Therefore, 
\be
\sum_i \vert \bra e_i \vert f_{-i} \ket \vert^2 =
\sum_i \vert \bra e_i \vert f_{1-i} \ket \vert^2 =
\sum_i \vert \bra e_i \vert f_{2-i} \ket \vert^2
\ee
Combining with the normalization conditions for any
$i,j$
\be
\sum_k \vert \bra e_k \vert f_j \ket \vert^2 = \sum_k
\vert \bra e_i \vert f_k \ket \vert^2 = 1
\ee
gives a linear system of 7 independent equations in
the 9
unknowns $\vert \bra e_i \vert f_j \ket \vert^2$ whose
solutions look like this:
\be
Ê(\vert \bra e_i \vert f_j \ket \vert^2)_{ij} =
\pmatrix{\vert a \vert^2 & \vert c \vert^2 & \vert b
\vert^2 \cr \vert b \vert^2 & \vert a \vert^2 & \vert
c \vert^2 \cr \vert c \vert^2 & \vert b \vert^2 &
\vert a \vert^2}
Ê\ee
ÊThat is, the quantity $\vert \bra e_i
\vert f_j \ket \vert^2$ depends only on $(j-i)$ mod 3.
Ê
\item{} Let $V$ be the unitary matrix
whose $(i,j)$ entry is given by $\bra e_i \vert f_j
\ket$.Ê From above, $\vert V_{i,j} \vert$ depends only
on $(j-i)$ mod 3.
We would like to have $V_{i,j} $ itself depend only on
$(j-i)$ mod 3. We accomplish this by 
adjusting the phases of the $
\vert e_i \ket$ and $\vert f_j \ket$, which is
equivalent to finding diagonal unitaries 
$U_1$ and $U_2$ such that
\be\label{desiredform}
V' = U_1VU_2^\daggerÊ = \pmatrix{aÊ &Ê cÊ &Ê bÊ \cr 
bÊ &Ê a 
&Ê cÊ \cr 
cÊ &Ê b &Ê aÊÊ }
\ee
for some 
$a,b,c \in C$. Write $m_{ij} = \arg({\bra e_i \vert
f_j
\ket})$ and 
\bee
U_1 = \pmatrix{1 &0&0 \cr 0& e^{i\alpha} &0 \cr 0&0&
e^{i\beta}} \quad U_2 = \pmatrix{1 &0&0 \cr0 &
e^{i\gamma} &
0\cr0 &0& e^{i\delta}} 
\eee
Solving a system of 3 linear equations in the phases
of the first two 
columns of $V$ allows us to set:
\be
\gamma & = & \frac{1}{3} \sum_{j = 0}^2
(m_{1j}-m_{0j}) \\
\alpha & = & m_{00} - m_{11} + \gamma \\
\beta & = & m_{01} - m_{20} - \gamma 
\ee
Put these values into $U_1$ and $U_2$ and choose
$\delta$ to adjust the top right corner, whichÊ gets
our matrix into the form
\be
V' =\pmatrix{aÊ &Ê cÊ &Ê bÊ \crÊ bÊ &Ê a 
&Ê ce^{i\delta_1}Ê \cr 
cÊ &Ê b &Ê ae^{i\delta_2}ÊÊ }
\ee
The fact that $V'$ is unitary implies its columns are
orthogonal, yielding the three equations
\be
\pmatrix{1 & 1 & 1 \cr e^{i\delta_1} & 1 &
e^{i\delta_2} \crÊ e^{-i\delta_2} & e^{-i\delta_1} &
1} \pmatrix{\overline{a}c \cr \overline{c}b \cr
\overline{b}a} = \pmatrix{0 \cr 0\cr 0}
\ee
The determinant of the above matrix cannot be zero unless 
$e^{i\delta_1} = e^{i\delta_2} = 1$, which means that in fact $V'$
is already in the desired formÊ (\ref{desiredform}).

Adjusting our matrix $V$ was equivalent to adjusting
the phases of 
the vectors $\vert e_i \ket$ and $\vert f_j \ket$.
Therefore, without loss of generality, we 
assume that $\bra e_i \vertÊ f_j \ket$
depends only on $(j-i)$ mod 3 and define
\be
A_{k} :=Ê \bra e_i \vertÊ f_{k+i} \ket 
\ee
which is independent of $i$.
\item{}Ê For $x \in \lbrace 0,1,2 \rbrace$ define:
\be
U\vert x \ket = \frac{1}{\sqrt{3}}\sum_{i=0}^2 \omega^{ix} 
\vert e_i \ket
\ee
Explicit
calculation shows that for all $x$, the vectors
$B_1U\vert x \ket, B_2U\vert x \ket, B_3U\vert x \ket$
are pairwise orthogonal:
\be
3\bra x \vert U^\daggerÊ B_2^\daggerÊ B_1 U \vert x
\ket &=& \sum_k
\omega^{-kx}Ê \omega^k 
\omega^{kx}= 0
\ee
\be
3\bra x \vert U^\daggerÊ B_3^\daggerÊ B_2 U \vert x
\ket &=&
\sum_{k,i,l} \omega^{-kx} \omega^i \omega^{lx}
\bra e_k \vert 
f_i \ket \bra f_i \vert e_l \ket \\ 
& = & \sum_{k,i,l} \omega^{(l-k)x} \omega^i A_{i-k}
\overline{A_{i-l}}Ê \\
&=&Ê \sum_{k',l'} \left. (\omega^{(k'-l')x} 
A_{k'} \overline{A_{l'}}Ê ) \sum_i \omega^i \right. =
0
\ee
\be
3\bra x \vert U^\daggerÊ B_3^\daggerÊ B_1 U \vert x
\ket &=& \bra x
\vert U^\daggerÊ B_3^\daggerÊ B_2 B_2^\daggerÊ B_1 U
\vert x \ket \\
& = & \sum_{k,i,l} \omega^{(l-k)x} \omega^{i+l}
A_{i-k} \overline{A_{i-l}}Ê \\
&=&Ê \sum_{k',l'} \left. (\omega^{(k'-l')x}\omega^{
- l'} A_{k'} \overline{A_{l'}}Ê \right) \sum_i
\omega^{2i}
= 0
\ee
\end{enumerate}
This proves that for all $x$, the vectors
$B_1U\vert x \ket, B_2U\vert x \ket, B_3U\vert x \ket$
are orthogonal and hence can be perfectly
distinguished. 
Ê
\subsection{Proof of Proposition \ref{CUB}}
Let ${\cal{B}} = \lbrace \vert b_1 \ket,Ê \ldots \vert
b_n \ket \rbrace$ be the common unbiased basis. We
need to show that for any $i \ne j$ and any $k$, the
vectors $B_i\vert b_k \ket$ and $B_j\vert b_k \ket$
are orthogonal. Using the eigenbasis ${\cal{A}}
_{ij}$, write
\be
B_i^\dagger B_j = \sum_s \lambda_s \vert e_s \kb e_s
\vert
\ee
Then for all $k$,
\be
\bra b_k \vert B_i^\dagger B_j \vert b_k \ket & = &
\sum_s 
\lambda_s \vert \bra b_k \vert e_s \ket \vert^2 \\
& = & \frac{1}{n} \sum_sÊ \lambda_s \\
& = & \frac{1}{n} \tr B_i^\dagger B_j = 0
\ee
\section{Proofs on the Worst Cases for Distinguishing
Orthogonal States}\label{imperfect}
Throughout what follows, let $V$ be the true identity
of the vector $\vert \Psi_i \ket$, and let $Z$ be
Alice and Bob's best guess of the the value of $V$,
which we assume is also the outcome of the final
measurement. Their goal, then, is to maximize $P(Z =
V)$.
\subsection{Proof of Propositions \ref{fmeprops} and
\ref{fprops} Using the Lemmas }
Setting $j = 2$ in 
Lemma \ref{propertieslemma} gives us the desired lower
bounds, since $f_{me}(2,n) = f(2,n) = 1$.
As long as $k
\le n^2$, there exist $k$ orthogonal maximally
entangled vectors in $C^n \otimes C^n$, so Lemma 
\ref{upperboundlemma} implies that $f_{me}(k,n)
\le
\frac{n}{k}$.
In the case $k \le 9$, we know $f_{me}(3,3) =
1$ so
\be
\frac{3}{k} = \frac{3}{k} f_{me}(3,3) \le f_{me}(k,3)
\le 
\frac{3}{k}
\ee
so $f_{me}(k,3) = \frac{3}{k}$.
Similarly, ifÊ $k \le m^2 \le n^2$, then $f(k,n) \le
f_{me}(k,m)
$, since we can
embedÊ maximally entangled $C^{m} \otimes C^{m}$
vectors
into $C^n \otimes C^n$. The minimum value of $m$ for
which we can do this is $\lceil \sqrt{k} \rceil$,
which implies 
\be
f(k,n) \le \frac{\lceil \sqrt{k} \rceil}{k}
\label{rootkbound}
\ee
In the case $2 \le k \le 4$, $\lceil \sqrt{k} \rceil = 2$ and
\be
\frac{2}{k} = \frac{2}{k} f(2,n) \le f(k,n) \le 
\frac{2}{k}
\ee
which implies
\be
f(k,n) = \frac{2}{k}
\ee
Ê
\subsection{Proof of Lemma \ref{propertieslemma}}
We prove the lemma for the function $f(k,n)$; the
proof for $f_{me}$ is identical.
Given any orthogonal vectors $\vert \Psi_i \ket \in
\lbrace \vert \Psi_1
\ket, \ldots, \vert \Psi_k \ket \rbrace$ with
probabilities $p_1 \ge p_2 \ge \ldots \ge p_k$. There
exists an algorithm that can distinguish the first $j$
of these vectors with probability at least $f(j,n)$.
Applying this algorithm to the received vector $\vert
\Psi_i \ket$ cannot succeedÊ if $i > j$, but clearly:
\be
P(Z = V) & \ge & P(Z = V,Ê i \le j) \\
& = & P(i \le j) P(Z = V \vert i \le j)Ê \\
& \ge & \frac{j}{k} f(j,n)
\ee
which gives the desired lower bound on $f(k,n)$.

\subsection{Proof of Lemma \ref{upperboundlemma}}
For this proof, we will need to examine the
measurement process more closely. As mentioned
earlier, we will assume 
that the protocol terminates with probability 1. In
fact, through the 
calculation, we will assume there exists an $r$ such
that the protocol terminates after at most $r$ rounds
of
communication. 
Completing the argument for arbitrary $r$ is
sufficient. Let $R$ be the actual number of rounds
needed to complete to protocol and let $p_r$ be the
probability that more than $r$ rounds are needed. Then
\be
P(Z=V) & = & (1-p_r)P(Z = V \vert R \le r) +Ê p_rP(Z =
V \vert R > r) \\
& \le & P(Z = V \vert R \le r) + p_r
\ee
Our proof will show that for any $r$, $P(Z = V \vert R
\le r) \le \frac{n}{k}$. Taking the limit as $r
\rightarrow \infty$, $ p_r$  gets arbitrarily small and we can bound $P(Z=V)$ by
$\frac{n}{k}$. 

The actions of Alice and Bob will consist of adding
ancilla systems, 
performing unitary operations, and performing
measurements. All of 
these can be encoded into a POVM. Alice measures
first; we write her POVM as ${\cal{X}}^T =\lbrace
X_1^T, X_2^T, \ldots X_{k_1}^T \rbrace $.Ê (Because we will eventually apply Lemma \ref{transposelemma} to show the effect of Alice's POVM on Bob's system, we write it in terms of the transpose.) Suppose Alice
gets the result $j_1$; then Bob uses a POVM that
depends on $j_1$: $ {\cal{E}}_{j_1}= \lbrace E(j_1)_1,
E(j_1)_2 \ldots E(j_1)_{k_2}\rbrace $. Alice then measures
in a POVM that depends on $j_1$ and $j_2$, and so on.
After $r$ rounds of measurement,
Alice and Bob have effectively measured using the POVM
\be
\lbrace {\cal{X}}_{j_1,j_2,\ldots j_{r-1}}^T \otimes
{\cal{E}}_{j_1,j_2,\ldots j_r} : j_1, j_2, \ldots, j_r
\ge 0 \rbrace
\ee
which are defined recursively as inÊ \cite{QNWE}:
\be 
{\cal{X}}_{j_1,j_2,\ldots j_{r-1}}^T &=&
X^T(j_1,j_2,\ldots,
j_{r-2})_{j_{r-1}}{\cal{X}}_{j_1,j_2,\ldots j_{r-3}}
^T \nonumber \\
{\cal{E}}_{j_1,j_2,\ldots j_r} &=&E(j_1,j_2,\ldots,
j_{r-1})_{j_r} {\cal{E}}_{j_1,j_2,\ldots j_{r-2}}
\label{sequencemeasures}
\ee
The subscripts show that each measurement
depends on the previous outcomes. Here eachÊ $X^T(m_0)$ and  $E(m_1)$
 is a POVM, where $m_0$ is a vector encoding an even number of previous outcomes and $m_1$ encodes an odd number. This corresponds to the fact that Alice and Bob alternate measurements, so Alice's action will always depend on an even number of previous results while's Bob's will always depend on an odd number. As usual, we have the normalization 
\be
Ê\sum_i (X(m_0)_i^T)^\dagger X(m_0)_i^T & =&
ÊI_{d_A(m_0)} = \sum_i X(m_0)_i X(m_0)_i^\dagger \\
\sum_{i}E(m_1)_i^\dagger
E(m_1)_i &=& I_{d_B(m_1)} 
 \ee
where $d_A(m_0)$ and $d_B(m_1)$ are sufficiently large
dimensions to include any ancilla spaces. 
 
 Alice and
Bob start with the state $\vert \Psi_i \ket = (I
\otimes B_i)\vert ME_n \ket$ and then apply the POVM
above, getting results $m = (j_1,j_2,\ldots j_{r-1})$ and $j_r$, for $r$ an
even number.Ê Then, using Lemma
\ref{transposelemma}, their state now looks like
\be
({\cal{X}}_m^T \otimes
{\cal{E}}_{m,j_r}) (I \otimes B_i)\vert
ME_n \ket = I \otimes (Ê {\cal{E}}_{m,j_r} B_i{\cal{X}}_m) \vert ME_n
\ket
\ee
This state is not normalized--its magnitude indicates
the probability of this outcome. Without loss of
generality, we assume that the final measurement
identifies the best guess of the value of $V$. This
gives us a more 
formal definition of our optimal
measurement, where we sum over all outcomes with the final output equal to the correct state identity: 
\be
P(\lbrace \vert \Psi_i \ket \rbrace, \lbrace
p_i \rbrace) &:=& \sup_{{\cal{X}},{\cal{E}}} P(Z = V) \\
 P(Z = V) &= &Ê \sum_i P(Z = V=i) \\
& = &ÊÊ  \sum_{i,m} p_i \bra
\Psi_i \vert (\overline{ {\cal{X}}}_{m}
{\cal{X}}_{m}^T \otimes {\cal{E}}_{m,i}^\dagger
{\cal{E}}_{m,i} \vert \Psi_i \ket \\
 & = &ÊÊ \sum_{i,m} p_i \bra
ME_n \vert (I \otimes {\cal{X}}_m^\dagger B_i^\dagger {\cal{E}}_{m,i}^\dagger
{\cal{E}}_{m,i} B_i  {\cal{X}}_m \vert ME_n \ket \\
&=& \frac{1}{n}
\sum_{i,m} p_i
\tr (X_m^\dagger B_i^\dagger{\cal{E}}^\dagger _{m,i}{\cal{E}}_{m,i}
B_iX_m) \label{totalprob}
\ee
 
The measurements might make use
of ancilla systems, so we write $P_A$ and $P_B$Ê as the
projections back
onto our original Alice and Bob spaces; since each $B_i$ maps Alice's space to Bob's, we see that $P_BB_i = B_iP_A= B_i$. Recall also that $\tr
B_i^\dagger B_i = n$
by assumption. 

We may now turn to the lemma, which assumes that
$\vert \Psi_i \ket = (I \otimes U_iB)\vert ME_n \ket$
with $U_i$ unitary and $B$ fixed.
Suppose Alice and Bob make $r$
measurements with the POVMsÊ described in
(\ref{sequencemeasures}). We assume that $r$  is even so Bob measures last--we can always append a trivial measurement to make this so. 
Suppose that the first $r-2$ measurement outcomes are
contained in the vector $m = (j_1, j_2, \ldots,
j_{r-2})$. For simplicity we write $j_{r-1}$ as $j$ and assume that $Z=V$ if and only if Bob's final measurement $j_r
= i$. Plugging this into (\ref{totalprob}) and setting $p_i = \frac{1}{k}$ yields
\be
P(Z = V) 
& = & \frac{1}{kn} \sum_{m, j,i} \tr (X_m^\dagger B_i^\dagger{\cal{E}}^\dagger _{m,i}{\cal{E}}_{m,i}
B_iX_m) \\
& = & \frac{1}{kn} \sum_{m, j,i} \tr (X_m^\dagger B_i^\dagger P_B{\cal{E}}^\dagger _{m,i}{\cal{E}}_{m,i}
P_B B_iX_m) \\
& \le & \frac{1}{kn} \sum_{m, j,i} (\tr P_B
{\cal{E}}_{m, j,i}^\dagger {\cal{E}}_{m, j,i} P_B
)(\tr B_i
{\cal{X}}_{m,j} {\cal{X}}_{m,j} ^\daggerÊ B_i^\dagger 
) 
\label{Holderhere}\\
& = & \frac{1}{kn} \sum_{m, j,i} (\tr P_B
{\cal{E}}_{m,
j,i}^\dagger {\cal{E}}_{m, j,i} )(\tr B^\dagger B
{\cal{X}}_{m,j}
{\cal{X}}_{m,j} ^\daggerÊ )Ê \label{pause}
\ee
In (\ref{Holderhere}), we use the fact that for 
matrices $A,B \ge 0$, $\tr AB \le (\tr A)(\tr B)$, and in (\ref{pause}) we use the assumption of the lemma that $B_i = U_iB$. The
key observation now is that there is no $i$ in the
second term of (\ref{pause}); rewriting the first term
as in (\ref{sequencemeasures}) shows that summing the
first term over $i$
yields the identity matrix on the inside,
allowing us to drop two subscripts, not just one:
\be
\tr \left(\sum_i P_B {\cal{E}}_{m, j,i}^\dagger
{\cal{E}}_{m, j,i}
\right)
& = & \tr \left(\sum_i P_BÊ {\cal{E}}_m^\dagger
E(m,j)_i^\dagger E(m,j)_i {\cal{E}}_m\right) \\
& = &\tr (P_B {\cal{E}}_m^\daggerÊ {\cal{E}}_m )
\ee
This corresponds to the fact that Alice does nothing during Bob's measurement phase. We now have
\be
P(Z = V) & \le & \frac{1}{kn} \sum_{m, j} (\tr P_B
{\cal{E}}_m^\daggerÊ 
{\cal{E}}_m )
(\tr B^\dagger B {\cal{X}}_{m,j}
{\cal{X}}_{m,j} ^\daggerÊ )Ê 
\ee
Now, there is no $j$ in the
first term, only in the second, so we can likewise sum
to get the 
identity on the inner term. Alternating in
this way, we can count back through the measurements
until
they all sum to the identity and we are left with
\be\label{laststep}
P(Z = V) 
& \le & \frac{1}{kn} \tr(P_B) \tr(B^\dagger B ) = 
\frac{1}{kn} (n)(n) = \frac{n}{k}
\ee

This shows that even if Alice and Bob add ancilla
systems to do their measurements, the relevant bound
comes from the dimension of Bob's system. This proves the 
lemma.
\subsection{Proof of Proposition \ref{schmidtprop}}
In equation (\ref{Holderhere}), we insert the projection onto Alice's space $P_A$ and use H\"older's
Inequality to note that
\be
\tr B_i {\cal{X}}_{m,j} {\cal{X}}_{m,j} ^\dagger 
B_i^\daggerÊ & = &
\tr B_i 
P_A{\cal{X}}_{m,j} {\cal{X}}_{m,j} ^\daggerÊ P_A
B_i^\daggerÊÊ \\
&\le&
\nrm B_i^\dagger B_i \nrm_\inftyÊ \trÊ P_A
{\cal{X}}_{m,j}
{\cal{X}}_{m,j} ^\dagger P_A \\
& \le & n \lambda_M \trÊ P_A{\cal{X}}_{m,j}
\label{lambdasub}
{\cal{X}}_{m,j} ^\daggerÊ P_A
\ee
since $\nrm B_i^\dagger B_i \nrm_\infty \le \max_i
\nrm
B_i^\dagger B_i \nrm_\infty = 
n\lambda_M$.
Aside from the new factor of $n
\lambda_M$, the rest of the calculation from Lemma
\ref{upperboundlemma} remains unchanged, inserting
$P_A$
for $B$ so that (\ref{laststep}) becomes
\be
P(Z = V) 
& \le & \frac{1}{kn} (n\lambda_M) \tr(P_B) \tr(P_A ) = 
\frac{\lambda_M}{k} (n)(m) = \frac{\lambda_M mn}{k}
\ee

Ê
\subsection{Proof of Proposition \ref{infoprop}}
The lower bound comes from the idea of tossing out all
but two of the vectors and distinguishing them
perfectly. At worst, this process gives you
$\frac{2}{k}\log{2}$ bits of information. 
Ê
The upper bound arises in the case of $k$ states to
which Lemma \ref{upperboundlemma} applies. The joint
probability distribution on $(V,Y,Z)$ must have two
properties. First, that the marginal distribution on
$V$ is uniform, since the states are equally likely.
Second, by relabeling in Lemma \ref{upperboundlemma},
we see that for any permutation $\sigma \in S_k, P(Z =
\sigma(V)) \le \frac{n}{k}$. The set of distributions
with these properties is a convex set on which the
mutual information is convex. The extreme points of
this set are distributions for which $Z$ takes on only
$n$ values and $Y$ is a
function of $Z$. Hence the maximum happens at an
extreme
point and 
\be
I(V; YZ) \le H(YZ) = H(Z) \le \log{n}
\ee
This implies that the maximum mutual information in
this case is $\log{n}$. Making $n$ as small as
possible, we see that
\be
g(k,n) \le \log{\lceil \sqrt{k} \rceil}
\ee

\section{Conclusion}
In summary, we have demonstrated that several classes
of maximally 
entangled states that can be distinguished using
LOCC. By examining the measurement process
itself, we have explored bounds on both the success
probability and the mutual information and shown that
the well-understood $C^2 
\otimes C^2$ Bell basis provides the worst case of $3$
or $4$ vectors with respect to either of these
measures. In the process, we have identified some sets
of states that cannot be
perfectly 
distinguished. It is hoped that through better
understanding best and worst cases of the
distinguishing problem, we can further our
understanding of the interplay between locality and
entanglement.
 
\medskip
\noindent{\bf{Acknowledgements:}} Many thanks to Chris
King for suggesting this line of inquiry, and to him
and Beth Ruskai for helpful discussions. The author
was supported in part by National Science Foundation
Grant DMS-0400426.
%\noindent{\bf{Appendix?:}} No.
{~~}
Ê\end{document}